\documentclass[a4paper,onecolumn]{article}
\usepackage{color}
\usepackage{graphicx}
\usepackage{amsfonts}
\usepackage{amssymb}
\usepackage[verbose,a4paper,tmargin=1.0in,bmargin=1.0in,lmargin=1.0in,rmargin=1.0in,nohead]{geometry}
\usepackage[super,sort&compress]{natbib}
\bibpunct{[}{]}{,}{s}{}{;} 

\newcommand{\mysection}[1]{\subsection*{\textsf{\normalsize#1}}}

\usepackage{setspace}
\usepackage{balance}

\newcommand{\Hop}{\hat{H}}
\newcommand{\sigmax}{\hat{\sigma}_x^{}}

\newcommand{\sigmaz}{\hat{\sigma}_z^{}}
\newcommand{\sigmap}{\hat{\sigma}_+^{}}
\newcommand{\sigmam}{\hat{\sigma}_-^{}}
\newcommand{\sigmapm}{\hat{\sigma}_\pm}
\newcommand{\aop}{\hat{a}}
\newcommand{\adop}{\hat{a}^\dag}

\newcommand{\aplus}{\adop+\aop}
\newcommand{\Uop}{\hat{U}}

\newcommand{\Sop}{\hat{S}}
\newcommand{\Sdop}{\hat{S}^\dag}
\newcommand{\Piop}{\hat{\Pi}}
\newcommand{\Nop}{\hat{N}}
\newcommand{\sigmaxs}{\hat{\sigma}_x^\star}

\newcommand{\sigmazs}{\hat{\sigma}_z^\star}

\newcommand{\kg}{\left|{\rm g}\right>}
\newcommand{\ke}{\left|{\rm e}\right>}
\newcommand{\bg}{\left<{\rm g}\right|}
\newcommand{\be}{\left<{\rm e}\right|}
\newcommand{\Ip}{I_{\rm p}}
\newcommand{\Phix}{\Phi_{\rm x}}

\newcommand{\Pe}{P_{\rm e}}
\newcommand{\omegaq}{\omega_{\rm q}}
\newcommand{\omegar}{\omega_{\rm r}}
\newcommand{\omegaf}{\omega_{\rm f}}
\newcommand{\Deltas}{\Delta^\star}
\newcommand{\Omegas}{\Omega^\star}
\newcommand{\epsilons}{\epsilon^\star}
\newcommand{\thetas}{\theta^\star}

\newcommand{\phimop}{\hat{\varphi}_{\rm m}}
\newcommand{\half}[1]{\frac{#1}{2}}

\newcommand{\expM}[1]{{\rm e}^{-i#1}}
\newcommand{\expP}[1]{{\rm e}^{+i#1}}

\begin{document}

\newlength{\myfigsize}
\setlength\myfigsize{0.5\textwidth}

\renewcommand{\topfraction}{1.0}
\renewcommand{\bottomfraction}{1.0}
\renewcommand{\textfraction}{0.0}

\noindent\parbox{\textwidth}{\flushleft\textsf{\Huge
Two-photon probe of the Jaynes-Cummings model and symmetry breaking in circuit QED}}
 \vspace{3mm}

\noindent\parbox{\textwidth}{\flushleft
\textsf{\Large
        Frank~Deppe$^{\mathsf{1,2,*,\dagger}}$,
        Matteo~Mariantoni$^{1,\dagger}$,
        E.~P.~Menzel$^{1}$,
        A.~Marx$^{1}$,
        S.~Saito$^{2}$,
        K.~Kakuyanagi$^{2}$,
        H.~Tanaka$^{2}$,
        T.~Meno$^{3}$,
        K.~Semba$^{2}$,
        H.~Takayanagi$^{4,5}$,
        E.~Solano$^{6,7}$
        and R.~Gross$^{1}$
       }
 \vspace{2mm}

\textsf{\textbf{\small $^{1}$Walther-Mei{\ss}ner-Institut, Bayerische Akademie der
Wissenschaften, Walther-Mei{\ss}ner-Str.~8, D-85748 Garching, Germany\\
$^{2}$NTT Basic Research Laboratories, NTT Corporation, Kanagawa, 243-0198, Japan\\
$^{3}$NTT Advanced Technology, NTT Corporation, Kanagawa, 243-0198, Japan\\
$^{4}$Tokyo University of Science, 1-3 Kagurazaka, Shinjuku, Tokyo, 162-8601, Japan\\
$^{5}$International Center for Materials Nanoarchitectronics, NIMS, Tsukuba 305-0003, Japan\\
$^{6}$Physics Department, ASC and CeNS,
Ludwig-Maximilians-Universit\"{a}t, Theresienstr.~37, 80333 M\"{u}nchen, Germany\\
$^{7}$Departamento de Qu\'{\i}mica-F\'{\i}sica, Universidad del Pa\'{\i}s Vasco - Euskal Herriko Unibertsitatea, Apdo. 644, 48080 Bilbao, Spain\\
$^{*}$e-mail: frank.deppe@wmi.badw-muenchen.de\\
$^{\dagger}$authors with equal contribution to this work
\vspace{5mm}
}}}

{\noindent\rule[2mm]{\textwidth}{0.1mm}}

 {\noindent\bf
Superconducting qubits\cite{Makhlin:2001a,Wendin:2006a} behave as artificial two-level atoms and are used to investigate fundamental quantum phenomena. In this context, the study of multi-photon excitations\cite{Nakamura:2001a,Oliver:2005a,Saito:2006a,Sillanpaa:2006a,Wilson:2007a} occupies a central role. Moreover, coupling superconducting qubits to on-chip microwave resonators has given rise to the field of circuit QED\cite{Wallraff:2004a,Chiorescu:2004a,Johansson:2006a,Houck:2007a,Astafiev:2007a,Sillanpaa:2007a,Majer:2007a,Wallraff:2007a}. In contrast to quantum-optical cavity QED\cite{Thompson:1992a,Mabuchi:2002a,Haroche:2006a,Walther:2006a}, circuit QED offers the tunability inherent to solid-state circuits. In this work, we report on the observation of key signatures of a two-photon driven Jaynes-Cummings model, which unveils the upconversion dynamics of a superconducting flux qubit\cite{Orlando:1999a} coupled to an on-chip resonator. Our experiment and theoretical analysis show clear evidence for the coexistence of one- and two-photon driven level anticrossings of the qubit-resonator system. This results from the symmetry breaking of the system Hamiltonian, when parity becomes a not well-defined property\cite{Liu:2005a}. Our study provides deep insight into the interplay of multiphoton processes and symmetries in a qubit-resonator system.
}

In cavity QED, a two-level atom interacts with the quantized modes of an optical or microwave cavity. The information on the coupled system is encoded both in the atom and in the cavity states. The latter can be accessed spectroscopically by measuring the transmission properties of the cavity\cite{Thompson:1992a}, whereas the former can be read out by suitable detectors\cite{Haroche:2006a,Walther:2006a}. In circuit QED, the solid-state counterpart of cavity QED, the first category of experiments was implemented by measuring the microwave radiation emitted by a resonator (acting as a cavity) strongly coupled to a charge qubit~\cite{Wallraff:2004a}. In a dual experiment, the state of a flux qubit was detected with a DC superconducting quantum interference device (SQUID) and vacuum Rabi oscillations were observed~\cite{Johansson:2006a}. More recently, both approaches have been exploited to extend the toolbox of quantum optics on a chip\cite{Blais:2004a,Houck:2007a,Astafiev:2007a,Sillanpaa:2007a,Majer:2007a,Wallraff:2007a}. Whereas all these experiments employ one-photon driving of the coupled qubit-resonator system, multi-photon studies are available only for sideband transitions~\cite{Wallraff:2007a} or bare qubits~\cite{Nakamura:2001a,Oliver:2005a,Saito:2006a,Sillanpaa:2006a,Wilson:2007a}. The experiments discussed in this work explore, to our knowledge for the first time, the physics of the two-photon driven Jaynes-Cummings dynamics in circuit QED. In this context, we show that the dispersive interaction between the qubit and the two-photon driving enables real level transitions. The nature of our experiment can be understood as an upconversion mechanism, which transforms the two-photon coherent driving into single photons of the Jaynes-Cummings dynamics. This process requires energy conservation and a not well-defined parity~\cite{Liu:2005a} of the interaction Hamiltonian due to the symmetry breaking of the qubit potential. Our experimental findings reveal that such symmetry breaking can be obtained either by choosing a suitable qubit operation point or by the presence of additional spurious fluctuators~\cite{Simmonds:2004a}.

\begin{figure}[t]
   \centering
   \includegraphics[width=\myfigsize]{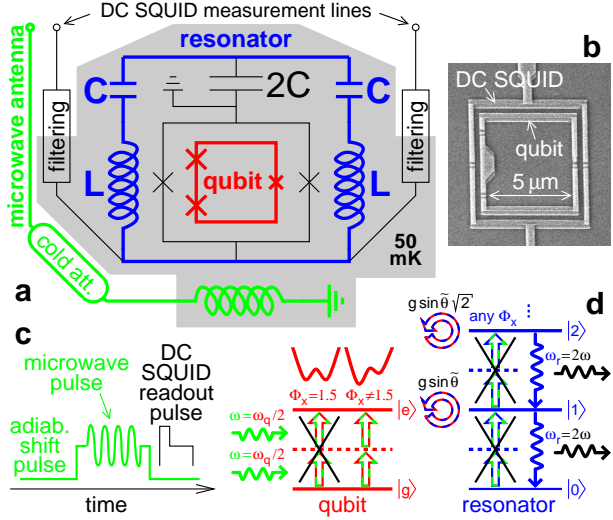}
   \caption{\footnotesize
{\bf Experimental architecture and theoretical model.}
{\bf a},
Schematic representation of the circuit. The Josephson junctions are represented by crosses. The readout DC~SQUID (black rectangle with two crosses), which is coupled to the three-Josephson-junction flux qubit (red) by a purely geometric mutual inductance, is shunted with an $LC$~circuit acting as a quantized resonator (blue). All circuit elements within the shaded area are at the base temperature $T\simeq50\,$mK of a dilution refrigerator. Microwave signals can be applied to qubit and resonator via an on-chip antenna (green). The signal-to-noise ratio of the driving is increased by means of cold attenuators.
{\bf b}, Scanning electron microscopy micrograph of the flux qubit and the readout DC~SQUID.
{\bf c}, Pulse scheme for qubit microwave spectroscopy with the adiabatic shift pulse method. First, the qubit is biased at a suitable readout point ($\Phix\ne1.5\Phi_0$) using a superconducting coil. Then, it is initialized in its ground state $\kg$ by waiting a sufficiently long time. In the next step, we use a rectangular pulse (``adiab.\ shift pulse'', green) to shift the qubit adiabatically to the desired operation point. There, it is irradiated with a $100\,$ns spectroscopy pulse (``microwave pulse'', green) of frequency $\omega/2\pi$. Both shift and spectroscopy pulse are applied via the microwave antenna. After the end of the shift pulse, a readout pulse to the DC~SQUID measurement lines probes the qubit state. By averaging over thousands of such measurements, the probability $\Pe$ to find the qubit in the excited state $\ke$ is calculated. Using the readout protocol described above, the qubit state can be detected via the DC~SQUID switching signal also at the optimal point, despite the fact that there the mean value $\Ip\left<\sigmaz\right>$ of the circulating current in the qubit loop vanishes\cite{Deppe:2007a}.
{\bf d},
Sketch of the upconversion dynamics describing the physics that governs the experiments discussed in this work, see Eq.~(\ref{eqn:twophotonhamiltonian}). The qubit (red) has a level splitting $\hbar\omegaq$ and the resonator (blue) frequency is $\omegar/2\pi$. We only consider the relevant case of two-photon driving (green) and assume that the system predominantly decays via the resonator. The two-photon driving frequency is $\omega$ and the qubit-resonator coupling strength is $g\sin\tilde{\theta}$, where $\sin\tilde{\theta}\equiv\Delta/\omegar\simeq0.63$. Depending on $\Phix$, the qubit potential (red double well; the $x$-axis represents the phase variable $\phimop$) is either symmetric or does not have a well-defined symmetry. Consequently, level transitions are allowed or forbidden respectively.
}
   \label{Fig01}
\end{figure}

The main elements of our setup, shown in Figs.~\ref{Fig01}a and b, are a three-Josephson-junction flux qubit, an $LC$-resonator, a DC~SQUID and a microwave antenna\cite{Kakuyanagi:2007a,Deppe:2007a}. The qubit is operated near the optimal flux bias $\Phix=1.5\Phi_0$ and can be described with the Hamiltonian $\Hop_{\rm q}=\left(\epsilon\sigmaz +\Delta\sigmax\right)/2$, where $\sigmax$ and $\sigmaz$ are Pauli operators.  From low-level microwave spectroscopy we estimate a qubit gap $\Delta/h=3.89\,$GHz. By changing $\Phix$, the quantity $\epsilon\equiv2\Ip\left(\Phix-1.5\Phi_0\right)$ and, in turn, the level splitting $\hbar\omega_{\rm q}\equiv\sqrt{\epsilon^2+\Delta^2}$ can be controlled. Here, $\pm \Ip$ are the clockwise and counterclockwise circulating persistent currents associated with the eigenstates $\left|\pm\right>$ of $\epsilon\sigmaz$. Far away from the optimal point, $\left|\pm\right>$ correspond to the eigenstates $\kg$ and $\ke$ of $\Hop_{\rm q}$. The qubit is inductively coupled to a lumped-element $LC$-resonator, which can be represented by a quantum harmonic oscillator, $\Hop_{\rm r}=\hbar\omega_{\rm r}\left(\adop\aop+1/2\right)$, with photon number states $\left|0\right>$, $\left|1\right>$, $\left|2\right>,\dots$ and boson creation and annihilation operators $\adop$ and $\aop$ respectively. This resonator is designed such that its fundamental frequency, $\omega_{\rm r}/2\pi=6.16\,$GHz, is largely detuned from $\Delta/h$. The qubit-resonator interaction Hamiltonian is $\Hop_{\rm q,r}=\hbar g\sigmaz\left(\aplus\right)$, where $g=2\pi\times115\,$MHz is the coupling strength. The $LC$-circuit also constitutes a crucial part of the electromagnetic environment of the readout DC~SQUID. In this way, the flux signal associated with the qubit states $\left|\pm\right>$ can be detected while maintaining reasonable coherence times and measurement fidelity\cite{Kakuyanagi:2007a,Deppe:2007a}.

\begin{figure}[t]
   \centering
   \includegraphics[width=\myfigsize]{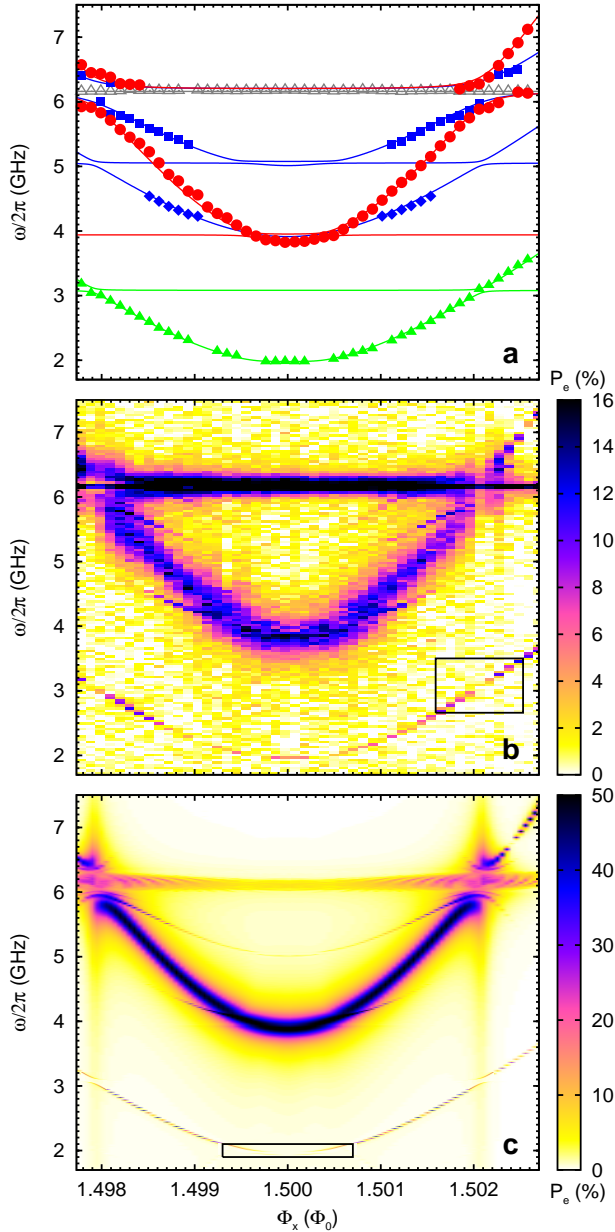}
   \caption{\footnotesize
{\bf Qubit microwave spectroscopy: Data and simulations.}
{\bf a},
Center frequency of the measured absorption peaks (symbols) plotted versus the flux bias: Red dots -- qubit one-photon signal; green triangles -- qubit two-photon signal; grey open triangles -- resonator; blue squares -- qubit-resonator blue sideband signal; blue diamonds -- qubit-fluctuator blue sideband signal. The lines are fits to the data based on the undriven Hamiltonian $\Hop_{\rm u}$. Of particular interest are the large and the small anticrossing for $\omega\approx\omegar$ and $2\omega\approx\omegar$ respectively. This constitutes direct evidence that the two-photon driving selectively excites only the qubit, but is strongly suppressed for the cavity. Consequently, the vacuum Rabi coupling $g$ can be obtained from two-photon spectroscopy. On the contrary, the one-photon driving populates the cavity and gives rise to the enhanced coupling $g\sqrt{\langle\Nop\rangle}$. 
{\bf b},
The measured probability $\Pe$ to find the qubit in the excited state plotted as a function of the flux bias and the microwave excitation frequency. The black box denotes the area shown in Fig.~\ref{Fig03}a.
{\bf c},
Probability $\Pe$ for the driven system described by $\Hop_{\rm d}$ obtained from numerical simulations using the time-trace-averaging method. The parameters are derived from the fit in {\bf a}. In our simulations, starting from the ground state $\kg$, the average over a full $100\,$ns time trace consisting of 10000 discrete time points already gives excellent agreement with the experimental data of {\bf b} (see main text). Since there are no terms describing dissipation in $\Hop_{\rm d}$, the linewidth of the peaks is caused by power broadening. The values of the mutual inductances are based on numerical estimates. The black box denotes the area shown in Fig.~\ref{Fig04}.
}
   \label{Fig02}
\end{figure}

In order to probe the properties of our system, we perform qubit microwave spectroscopy using an adiabatic shift pulse technique\cite{Kakuyanagi:2007a,Deppe:2007a} (Fig.~\ref{Fig01}c). The main results are shown in Figs.~\ref{Fig02}a and b. First, there is a flux-independent feature at approximately $6\,$GHz due to the resonator. Second, we observe two hyperbolas with minima near $4\,\textrm{GHz}\simeq\Delta/h$ and $2\,\textrm{GHz}\simeq\Delta/2h$, one with a broad and the other with a narrow linewidth. They correspond to the one-photon ($\omega=\omega_{\rm q}$) and two-photon ($2\omega=\omega_{\rm q}$) resonance condition between the qubit and the external microwave field. Additionally, the signatures of two-photon driven blue sideband transitions are partially visible. One can be attributed to the resonator, $\left|g,0\right>\rightarrow\left|e,1\right>$, and the other to a spurious fluctuator\cite{Simmonds:2004a}. We assume that the latter is represented by the flux-independent Hamiltonian $\Hop_{\rm f}=\left(\epsilons\sigmazs+\Deltas\sigmaxs\right)/2$ and coupled to the qubit via $\Hop_{\rm q,f}=\hbar g^\star\sigmaz\sigmazs$, where $\sigmaxs$ and $\sigmazs$ are Pauli operators. Exploiting the different response of the system in the anticrossing region under one- and two-photon driving, as explained in Fig.~\ref{Fig02}a, the center frequencies of the spectroscopic peaks can be accurately fitted to the undriven Hamiltonian $\Hop_{\rm u}=\Hop_{\rm q}+\Hop_{\rm r}+\Hop_{\rm f}+\Hop_{\rm q,r}+\Hop_{\rm q,f}$. Setting $\epsilons=0$ (cf.\ methods), we obtain $g/2\pi=115\,$MHz, $\langle\Nop\rangle\simeq10$, $\Ip=367\,$nA, $\omega_{\rm f}/2\pi\equiv\sqrt{{\epsilons}^2 +{\Deltas}^2}/h=3.94\,$GHz and $g^\star\sin\thetas=37\,$MHz, where $\sin\thetas\equiv\Deltas/\hbar\omegaf$. 

Further insight into our experimental results can be gained by numerical spectroscopy simulations based on the driven Hamiltonian $\Hop_{\rm d} = \Hop_{\rm u}+\Hop_{\rm m,q}+\Hop_{\rm m,r}+\Hop_{\rm m,f}$. Here, $\Hop_{\rm m,q}=\frac{\Omega}{2}\sigmaz\cos\omega t$, $\Hop_{\rm m,r}=\eta\left(\aplus\right)\cos\omega t$ and $\Hop_{\rm m,f}=\frac{\Omegas}{2}\sigmazs\cos\omega t$ represent the driving of the qubit, resonator and fluctuator respectively. We approximate the steady state with the time average of the probability $\Pe$ to find the qubit in $\ke$ (time-trace-averaging method). Inspecting Fig.~\ref{Fig02}c, we find that for the driving strengths $\Omega/h=244\,$MHz, $\eta/h=655\,$MHz and $\Omegas=0$ our simulations match well all the experimental features discussed above. Using $\eta$ and the relation $\langle\Nop\rangle=(\eta/\kappa)^2$ for the steady-state mean number of photons of a driven dissipative cavity, we estimate a cavity decay rate $\kappa\simeq210\,$MHz. This result is of the same order as $\kappa\simeq400\,$MHz estimated directly from the experimental linewidth of the resonator peak. The large $\kappa$ is due to the galvanic connection of the resonator to the DC~SQUID measurement lines (Fig.~\ref{Fig01}a).

\begin{figure}[t]
   \centering
   \includegraphics[width=\myfigsize]{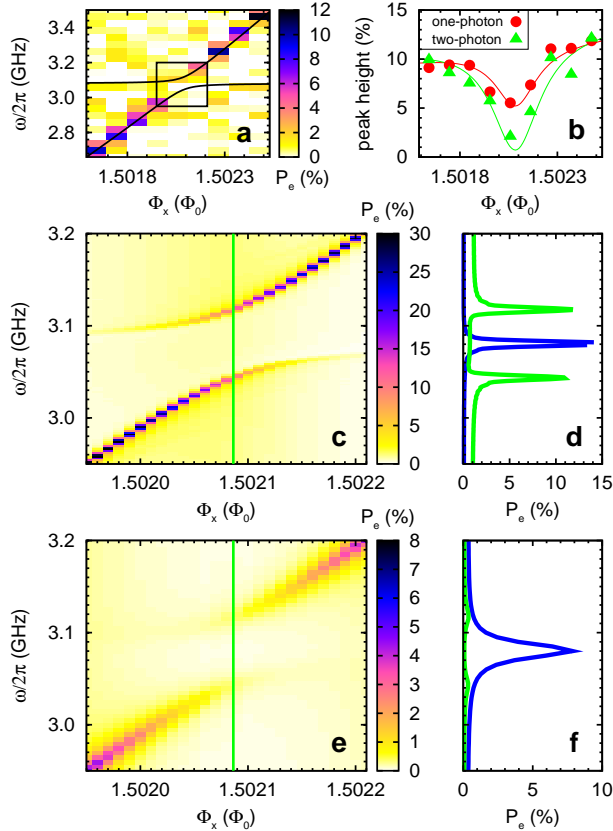}
   \caption{\footnotesize
{\bf Qubit microwave spectroscopy close to the qubit-resonator anticrossing under two-photon driving.}
{\bf a}, 
Measured probability $\Pe$ to find the qubit in the excited state plotted vs.\ the applied flux and the microwave excitation frequency. The black rectangle denotes the area shown in {\bf c} and {\bf e}. The solid lines represent the fit obtained by the undriven Hamiltonian $\Hop_{\rm u}$.
{\bf b}, 
Maximum height of the spectroscopy peaks under one-photon (red dots) and two-photon (green triangles) driving plotted vs. the flux bias. The solid lines are guides to the eye.
{\bf c},
Probability $\Pe$ obtained from simulations including driving, but no dissipation (time-trace-averaging method). The parameters are derived from the numerical fit of Fig.~\ref{Fig02}a. The signature of an anticrossing is visible.
{\bf d}, The green curve shows the split-peak profile of $\Pe$ along the vertical line in {\bf c}. The blue line shows the single-peak result obtained for the same flux bias, when turning off the qubit-resonator coupling by setting $g=0$.
{\bf e},
Probability $\Pe$ obtained from simulations including driving and dissipation based on the Lindblad formalism\cite{Blais:2004a}. For simplicity, the spurious fluctuator is not included here. We assume a qubit relaxation time $T_1=300\,$ns, a dephasing time $T_\varphi=15\,$ns and a resonator quality factor $Q\equiv\omegar/\kappa=2\pi\times6.16\,{\rm GHz}/400\,{\rm MHz}\simeq100$ corresponding to a resonator decay time $\tau\simeq3\,$ns. The spectroscopy signal is chosen to be the average over the last $20\,$ns of a $100\,$ns time trace. When the qubit and the resonator become degenerate, the spectroscopy signal fades away.
{\bf f},
Again, the green curve shows the split-peak profile of $\Pe$ along the vertical line in {\bf e}. The blue line shows the single-peak result obtained for the same flux bias when setting $g=0$. Differently from the nondissipative case ({\bf c} and {\bf d}), the single peak is approximately ten times higher than either one of the strongly reduced split peaks (green curve). This demonstrates that the vanishing of the spectroscopy signal observed in the experimental data (cf.\ {\bf a}, {\bf b}) and in the simulations with dissipation (cf.\ {\bf e}) is not caused by qubit decoherence.
}
  \label{Fig03}
\end{figure}

To elucidate the two-photon driving physics of the qubit-resonator system we consider the spectroscopy data near the corresponding anticrossing shown in Fig.~\ref{Fig03}a. For $2\omega=\omegaq=\omegar$, the split peaks cannot be observed directly because the spectroscopy signal is decreased below the noise floor $\delta\Pe\simeq1-2\%$. This results from the fact that the resonator cannot absorb a two-photon driving and its excitation energy is rapidly lost to the environment ($\kappa>g/2\pi$). In contrast, for the one-photon case ($\omega=\omegaq=\omegar$), there is a driving-induced steady-state population of $\langle\Nop\rangle\simeq10$ photons in the cavity.  Accordingly, the one-photon peak height shows a reduction by a factor of approximately two, whereas the two-photon peak almost vanishes, see Fig.~\ref{Fig03}b. To prove that this effect is only due to the resonator, we compare the simulation results from the time-trace-averaging method to those obtained with the standard Lindblad dissipative bath approach (Figs.~\ref{Fig03}c-f). Altogether, our experimental data and numerical simulations constitute clear evidence for the presence of a qubit-resonator anticrossing under two-photon driving.

The second-order effective Hamiltonian under two-photon driving can be derived using a Dyson-series approach (cf.\ methods). Starting from the first-order driven Hamiltonian $\Hop_{\rm d}$ and neglecting the cavity driving and the fluctuator because of large-detuning conditions, we obtain
\begin{eqnarray}
   \Hop^{(2)}_{}
   & \!\!=\!\! & 
    \frac{\hbar\omegaq}{2}\sigmaz
    +\frac{\Omega^2}{4\Delta}\sin^2\theta\cos\theta
    \Big(\sigmap\expM{2\omega t}+\sigmam\expP{2\omega t}\Big)
   \nonumber\\
   & &
    {}-\,\hbar g\sin\theta\Big(\sigmap\aop+\sigmam\adop\Big)
    +\hbar\omegar\left(\adop\aop+\frac{1}{2}\right)
    \,,
   \label{eqn:twophotonhamiltonian}
\end{eqnarray}
where $\sigmap$ and $\sigmam$ are the qubit raising and lowering operators, $\sin\theta\equiv\Delta/\omegaq$ and $\cos\theta\equiv\epsilon/\omegaq$. The upconversion dynamics sketched in Fig.~\ref{Fig01}d is clearly described by Eq.~(\ref{eqn:twophotonhamiltonian}). The first two terms represent the qubit and its coherent two-photon driving with angular frequency $\omega$. The last two terms show the population transfer via the Jaynes-Cummings interaction to the resonator. The Jaynes-Cummings interaction in this form is valid only near the anticrossings ($\theta\approx\tilde{\theta}$, $\tilde{\theta}\equiv\frac{\Delta}{\omegar}\simeq0.63$; cf.\ methods). As discussed before, the resonator will then decay emitting radiation of angular frequency $2\omega$.

\begin{figure}[t]
   \centering
   \includegraphics[width=\myfigsize]{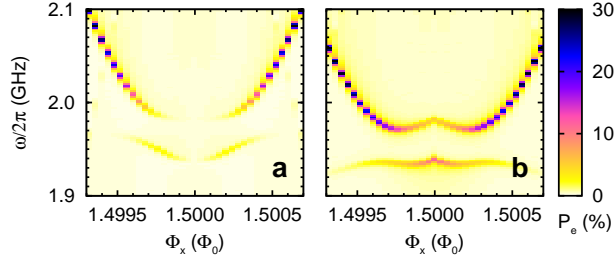}
   \caption{\footnotesize
{\bf Two-photon spectroscopy simulations close to the optimal point using the time-trace-averaging method.}
{\bf a},
Probability $\Pe$ plotted vs.\ driving frequency and flux bias. The parameters are the same as those used in Fig.~\ref{Fig02}c. In particular, the fluctuator parameters are $\epsilons=0\leftrightarrow\sin\thetas=1$ and $\Omegas=0$. The spectroscopy signal vanishes completely at the optimal point ($\Phix=1.5\Phi_0$) because of the specific selection rules associated with the symmetry properties of the Hamiltonian\cite{Liu:2005a}.
{\bf b},
Same as in {\bf a}, however, for $\sin\thetas=0.3$ and $\Omegas=280\,$MHz. Here, the coexistence of the flux-independent first-order $\sigmaxs$- and $\sigmazs$-terms of the fluctuator gives rise to a nonvanishing second-order $\sigmaxs$-term, even at the qubit optimal point. In other words, the presence of the fluctuator breaks the symmetry of the total system at the optimal point and, again, parity becomes not well defined. Consequently, the spectroscopy signal is partially revived. In reality, there is a high probability that not a single, but an ensemble of fluctuators with some distribution of frequencies and coupling strengths contributes to the symmetry breaking. Furthermore, when the experimental resolution is limited, a single peak will be detected and detailed structure such as the one of Fig~\ref{Fig04}b will not be visible. This is actually the case in our measurements (Fig.~\ref{Fig02}b).
}
   \label{Fig04}
\end{figure}

The model outlined above allows us to unveil the symmetry properties of our system. Even though the two-photon coherent driving is largely detuned, $\frac{\omegaq}{2}=\omega\gg\frac{\Omega}{2}\sin\theta$, a not well-defined symmetry of the qubit potential permits level transitions away from the optimal point. Because of energy conservation, i.e.\ frequency matching, these transitions are real and can be used to probe the qubit-resonator anticrossing. The effective two-photon qubit driving strength, $\left(\Omega^2\sin^2\theta/4\Delta\right)\cos\theta$, has the typical structure of a second-order dispersive interaction with the extra factor $\cos\theta$. The latter causes this coupling to disappear at the optimal point. There, the qubit potential is symmetric and the parity of the interaction operator is well defined. Consequently, selection rules similar to those governing electric dipole transitions hold\cite{Liu:2005a}. This is best understood in our analytical two-level model, where the first-order Hamiltonian for the driven diagonalized qubit becomes $\Hop^{(1)}_{\rm OP}=\frac{\Delta}{2}\sigmaz+\frac{\Omega}{4}\sigmax\left(\expP{\omega t}+\expM{\omega t}\right)$ at the optimal point. In this case, one-photon transitions are allowed because the driving couples to the qubit via the odd-parity operator $\sigmax$. In contrast, the two-photon driving effectively couples via the second-order Hamiltonian $\Hop^{(2)}_{\rm OP}=\frac{\Delta}{2}\sigmaz+\frac{\Omega^2}{16\Delta}\sigmaz\left(\expP{\omega t}+\expM{\omega t}\right)^2$. Since $\sigmaz$ is an even-parity operator, real level transitions are forbidden (cf.\ methods). We note that the second $\sigmaz$-term of $\Hop^{(2)}_{\rm OP}$ renormalizes the qubit transition frequency slightly and can be neglected in Eq.~(\ref{eqn:twophotonhamiltonian}), which describes the real level transitions corresponding to our spectroscopy peaks. The intimate nature of the symmetry breaking resides in the coexistence of $\sigmax$- and $\sigmaz$-operators in the first-order Hamiltonian $\Hop_{\rm d}$, which produces a nonvanishing $\sigmax$-term in the second-order Hamiltonian $\Hop^{(2)}_{}$  of Eq.~(\ref{eqn:twophotonhamiltonian}). This scenario can also be realized at the qubit optimal point by the fluctuator terms $\sigmaxs$ and $\sigmazs$. As illustrated in Fig.~\ref{Fig04}, their presence causes a revival of the two-photon signal and the discussed strict selection rules no longer apply. Accordingly, we observe only a reduction instead of a complete suppression of the two-photon peaks near the qubit optimal point in the experimental data of Fig.~\ref{Fig02}b.

In conclusion, we use two-photon qubit spectroscopy to study the interaction of a superconducting flux qubit with an $LC$-resonator. We show experimental evidence for the presence of an anticrossing under two-photon driving, permitting us to estimate the vacuum Rabi coupling. Our experiments and theoretical analysis shed new light on the fundamental symmetry properties of quantum circuits and the nonlinear dynamics inherent to circuit QED. This can be exploited in a wide range of applications such as parametric up-conversion, generation of microwave single photons on demand\cite{Mariantoni:2005a,Liu:2004a,Houck:2007a} or squeezing\cite{Moon:2005a}. 

\vspace{1mm}


\section*{\textsf{\large APPENDIX\\{\rule[5mm]{\columnwidth}{0.1mm}}\vspace*{-5mm}}}

{\it Two-photon driven Jaynes-Cummings model via Dyson series.} We now derive the effective second-order Hamiltonian describing the physics relevant for the analysis of the two-photon driven system. We start from the first-order Hamiltonian in the basis $\left|\pm\right>$,
\begin{eqnarray}
   \Hop & = &
      \half{\epsilon}\sigmaz
      +\half{\Delta}\sigmax
      +\hbar\omegar\left(\adop\aop+\half{1}\right)
      \nonumber\\
   &  &
      {}+\hbar g\,\sigmaz\left(\adop+\aop\right)
      +\half{\Omega}\sigmaz\cos\omega t\,.
   \label{eqn:hamiltonianinbarebasis}
\end{eqnarray}
Here, in comparison to $\Hop_{\rm d}$, the terms associated with the fluctuator are not included ($\epsilons=\Deltas=\Omegas=0$) because the important features are contained in the driven qubit-resonator system. Additionally, we focus on the two-photon resonance condition $\omegar=\omegaq=2\omega$. Thus, the driving angular frequency $\omega$ is largely detuned from $\omegar$ and the corresponding term in $\Hop_{\rm d}$ can be neglected ($\eta=0$). Next, we transform the qubit into its energy eigenframe and move to the interaction picture with respect to qubit and resonator, $\sigmapm\rightarrow\sigmapm{\rm e}^{\pm i\omegaq t}$, $\aop\rightarrow\aop\expM{\omegar t}$ and $\adop\rightarrow\adop\expP{\omegar t}$. After a rotating wave approximation we identify the expression $\Sdop\expP{\omega t}+\Sop\expM{\omega t}$, where the superoperator $\Sop\equiv\frac{\Omega}{4}\left(\cos\theta\,\sigmaz-\sin\theta\,\sigmam\right)$ and its Hermitian conjugate $\Sdop\equiv\frac{\Omega}{4}\left(\cos\theta\,\sigmaz-\sin\theta\,\sigmap\right)$. In our experiments the two-photon driving of the qubit is weak, i.e.\ the large-detuning condition $\omegaq-\omega=\omega\gg\frac{\Omega}{2}\sin\theta$ is fulfilled. In such a situation, it can be shown that the Dyson series for the evolution operator associated with the time-dependent Hamiltonian $-\hbar g\sin\theta(\sigmap\aop+\sigmam\adop)+(\Sop\expM{\omega t}+\Sdop\expP{\omega t})$ can be rewritten in an exponential form $\Uop=\expM{\Hop^{}_{\rm eff}t/\hbar}$, where
\begin{eqnarray}
   \Hop_{\rm eff}^{} & = & 
    {}-\hbar g\sin\theta\left(\sigmap\aop\expP{\delta t}
                              +\sigmam\adop\expM{\delta t}\right)
    +\frac{\big[\Sdop,\Sop\big]}{\hbar\omega}
   \nonumber\\
   & = &
    {}-\hbar g\sin\theta\left(\sigmap\aop\expP{\delta t}
                              +\sigmam\adop\expM{\delta t}\right)
   \nonumber\\
   & &
    {}+\frac{\Omega^2}{4\Delta}\left(\sin^2\theta\cos\theta\,\sigmax
    +\frac{1}{2}\sin^3\theta\,\sigmaz\right) \,.
   \label{eqn:twophotondrivingresult}
\end{eqnarray}
Here, $\delta\equiv\omegaq-\omegar$ is the qubit-resonator detuning. In Eq.~(\ref{eqn:twophotondrivingresult}), the dispersive shift $\frac{\Omega^2}{8\Delta}\sin^3\theta\,\sigmaz$ is a reminiscence of the full second-order $\sigmaz$-component of the interaction Hamiltonian, $\frac{\Omega^2}{16\Delta}\sin^3\theta\,\sigmaz\left(\expP{\omega t}+\expM{\omega t}\right)^2$. The terms proportional to $\sigmaz\exp^{\pm i2\omega t}$ are neglected implicitly by a rotating wave approximation when deriving the effective Hamiltonian $\Hop_{\rm eff}$ of Eq.~(\ref{eqn:twophotondrivingresult}). In this equation, the $\sigmaz$-term renormalizes the qubit transition frequency, and, in the vicinity of the anticrossing ($|\delta|\lesssim g\sin\tilde{\theta}$, $\sin\tilde{\theta}=\frac{\Delta}{\omegar}\simeq0.63$), the Hamiltonian $\Hop_{}^{(2)}$ of Eq.~(\ref{eqn:twophotonhamiltonian}) can be considered equivalent to $\Hop_{\rm eff}^{}$. In this situation, the symmetries of the system are broken and our experiments demonstrate the existence of real level transitions.

{\it Selection rules.} The potential of the three-Josephson-junction flux qubit can be reduced to a one-dimensional double well with respect to the phase variable $\phimop$\cite{Orlando:1999a}. At the optimal point ($\Phix=1.5\Phi_0$), this potential is a symmetric function of $\phimop$. For our experimental parameters, we can assume an effective two-level system. The two lowest energy eigenstates $\left|\rm g\right>$ and $\left|\rm e\right>$ are, respectively, symmetric and antisymmetric superpositions of $\left|+\right>$ and $\left|-\right>$. Thus, $\left|\rm g\right>$ has even parity and $\left|\rm e\right>$ is odd. In this situation, the parity operator $\Piop$ can be defined via the relations $\Piop\kg=+\kg$ and $\Piop\ke=-\ke$. The Hamiltonian of the classically driven qubit is $\frac{\Delta}{2}\sigmaz+\frac{\Omega}{2}\cos\omega t\,\sigmax$. For a one-photon driving, $\omega=\Delta/\hbar$ (energy conservation), the Hamiltonian in the interaction picture is $\frac{\Omega}{4}\sigmax$, where $\sigmax\equiv\kg\be+\ke\bg$. This is an odd-parity operator because the anticommutator $\{\Piop,\sigmax\}=0$ and, consequently, one-photon transitions are allowed. For a two-photon driving, $\omega=\Delta/2\hbar$ (energy conservation), the effective interaction Hamiltonian becomes $\frac{\Omega^2}{8\Delta}\sigmaz$, where $\sigmaz\equiv\ke\be-\kg\bg$. Since the commutator $[\Piop,\sigmaz]=0$, this is an even-parity operator and two-photon transitions are forbidden\cite{CohenTannoudji:1977a}. These selection rules are analogous to those governing electric dipole transitions in quantum optics. On the contrary, in circuit QED the qubit can be biased away from some optimal point. In this case, the symmetry is broken and the discussed selection rules do not hold. Instead, we find the finite transition matrix elements $\frac{\Omega}{4}\sin\theta$ and $\frac{\Omega^2}{4\Delta}\sin^2\theta\cos\theta$ for the one- and two-photon process respectively. Beyond the two-level approximation, the selection rules for a flux qubit at the optimal point are best understood by the observation that the double-well potential is symmetric there (Fig.~\ref{Fig01}d). Hence, the interaction operator of the one-photon driving is odd with respect to the phase variable ${\hat\varphi}_{\rm m}$ of the qubit potential\cite{Orlando:1999a,Liu:2005a}, whereas the one of the two-photon driving is even. Away from the optimal point ($\Phix\ne1.5\Phi_0$), the qubit potential has no well-defined symmetry and no selection rules apply.

{\it Spurious fluctuators.} The presence of spurious fluctuators in qubits based on Josephson junctions has already been reported previously\cite{Simmonds:2004a}. In principle, such fluctuators can be either resonators or two-level systems. Since our experimental data does not allow us to distinguish between these two cases, for simplicity, we assume a two-level system in the simulations. In the numerical fit shown in Fig.~\ref{Fig02}a, we choose $\epsilons=0$ due to the limited experimental resolution. Consequently, the coupling constant estimated from the undriven fit is not $g^\star$, but $g^\star\sin\thetas$. Away from the qubit optimal point, especially near the qubit-resonator anticrossings, the effect of the observed fluctuator can be neglected within the scope of this study. Near the optimal point, its effect on the symmetry properties of the system can be explained following similar arguments as given above for the flux qubit. However, it is important to note that, differently from $\sin\theta$ and $\cos\theta$, the fluctuator parameters $\sin\thetas$ and $\cos\thetas$ are constants, i.e.\ they do not depend on the quasi-static flux bias $\Phix$.


\mysection{Acknowledgements}

We thank H.~Christ for fruitful discussions. This work is supported by the Deutsche Forschungsgesellschaft through the Sonderforschungsbereich 631, the German Excellence Initiative via the Nanosystems Initiative Munich (NIM) and the EuroSQIP EU project. This work is partially supported by CREST-JST, JSPS-KAKENHI(18201018) and MEXT-KAKENHI(18001002).

\protect

\end{document}